\documentclass[preprint,showpacs]{revtex4}
\usepackage{mathrsfs}
\usepackage{amsfonts}

\usepackage{graphicx}
\usepackage{}% Include figure files
\usepackage{amsmath}% Include figure files
\usepackage{dcolumn}% Align table columns on decimal point
\usepackage{color}

\begin{document}
\title{Electromagnetically induced transparency and slow light in two-mode optomechanics}
\author{Cheng Jiang}
\email{chengjiang_phys@163.com}
\author{Hongxiang Liu}
\author{Yuanshun Cui}
\author{Xiaowei Li}
 \affiliation{School of Physics and Electronic Electrical Engineering, Huaiyin Normal University, 111 West Chang Jiang Road, Huaian 223001, China}
\author{Bin Chen}
\affiliation{College of Physics and Optoelectronics, Taiyuan University of Technology, Shanxi, Taiyuan 030024, China}
\date{\today}
\begin{abstract}
We theoretically demonstrate the mechanically mediated electromagnetically induced transparency in a two-mode cavity optomechanical system, where two cavity modes are coupled to a common mechanical resonator. When the two cavity modes are driven on their respective red sidebands by two pump beams, a transparency window appears in the probe transmission spectrum due to destructive interference. Under this situation the transmitted probe beam can be delayed as much as 4 $\mu$s, which can be easily controlled by the power of the pump beams. In addition, we also investigate the amplification of the transmitted probe beam owing to constructive interference when one cavity is driven on its blue sideband while another one is driven on its red sideband.
\end{abstract}
\pacs{42.50.Gy, 42.50.Wk, 42.50.Ex}

\maketitle
 \section{Introduction}
The emerging field of cavity optomechanics \cite{Kippenberg1,Kippenberg2,Marquardt,Aspelmeyer} studies the interaction between optical and mechanical modes via radiation pressure force, which enables to observe quantum mechanical behavior of macroscopic systems. Recent process in fabrication and cooling techniques paves the way towards realizing strong coupling at the single-photon level in optomechanical systems \cite{Brennecke,Rabl,Nunenkamp,Teufel1,Verhagen} and cooling the nanomechanical resonators to their quantum ground state \cite{Teufel2,Chan}. Moreover, the optical response of optomechanical systems is modified because of mechanical interactions, leading to the phenomenon of normal-mode splitting \cite{Groblacher} and electromagnetically induced transparency (EIT) \cite{Agarwal,Weis,Naeini,Karuza}. In EIT \cite{Fleischhauer} an opaque medium can be made transparent in the presence of a strong pump beam; the concomitant steep variation of the refractive index induces a drastic reduction in the group velocity of a probe beam, which can be used to slow and stop light \cite{Hau,Chang}. EIT has been first observed in atomic vapors \cite{Boller} and recently in various solid state systems such as quantum wells \cite{Phillips}, metamaterial \cite{Liu} and nitrogen-vacancy centers \cite{Santori}. In optomechanical systems, sub- and superluminal light propagation based on EIT have been observed both in optical \cite{Naeini,Karuza} and microwave domains \cite{Zhou}.

Most recently, two-mode optomechanics in which two optical modes are coupled to a mechanical mode have received a lot of research interest. Ludwig \emph{et al.} \cite{Ludwig} and K\'{o}m\'{a}r \emph{et al.} \cite{Komar} have theoretically shown that quantum nonlinearities can be enhanced significantly in two-mode optomechanical systems, which can be used in optomechanical quantum information processing with photons and phonons \cite{Stannigel}. Qu and Agarwal \cite{Qu} theoretically showed that double cavity optomechanical systems can be used both as memory elements as well as for the transduction of optical fields. Moreover, Hill \emph{et al.} \cite{Hill} and Dong \emph{et al.} \cite{Dong} have experimentally demonstrated coherent wavelength conversion of optical photons between two different optical wavelengths in optomechanical crystal nanocavity and silica resonator, respectively. In the present paper, we investigate the optical response of the two-mode optomechanical system in the simultaneous presence of two strong pump beams and a weak probe beam. When the two cavities are pumped on their red sidebands (i.e., one mechanical frequency, $\omega_m$, below cavity resonances, $\omega_1$ and $\omega_2$), respectively, a transparency window appears in the probe transmission spectrum. Furthermore, electromagnetically induced amplification has also been demonstrated when one cavity is pumped on its blue sideband while another one is pumped on its red sideband. The paper is organized as follows. Section \uppercase\expandafter{\romannumeral2} gives the theoretical model and method. Results and discussion are shown in Sec. \uppercase\expandafter{\romannumeral3}. A summary is presented in Sec. \uppercase\expandafter{\romannumeral4}.

\section{Model and Theory}
We consider an optomechanical system as shown in Fig. 1, where two optical cavity modes $a_k$($k=1, 2$) are coupled to a common mechanical mode $b$. The left cavity is driven by a strong pump beam $E_L$ with frequency $\omega_L$ and a weak probe beam $E_p$ with frequency $\omega_p$ simultaneously, and the right cavity is only driven by a strong pump beam $E_R$ with frequency $\omega_R$.
 In a rotating frame at
the pump frequency $\omega_{L}$ and $\omega_R$, the Hamiltonian of the two-mode optomechanical system reads as
follows \cite{Hill}: %{\setlength\arraycolsep{1pt}
%\begin{eqnarray}
%H=&&\sum_{k=1,2}\hbar\omega_{k}a_{k}^\dagger a_{k}+\hbar\omega_mb^\dagger b-\sum_{k=1,2}\hbar g_k
%a_k^\dagger
%a_k(b^\dagger+b)+i\hbar\sqrt{\kappa_{e,1}}(E_{L}a_{1}^\dagger e^{-i\omega_{L}t}-E_L^*a_{1}e^{i\omega_{L}t})\nonumber\\&&+i\hbar\sqrt{\kappa_{e,2}}(E_{R}a_{2}^\dagger e^{-i\omega_{R}t}-E_{R}^*a_{2}e^{i\omega_{R}t})
%+i\hbar\sqrt{\kappa_{e,1}}(E_{p}a_{1}^\dagger
%e^{-i\omega_{p}t}-E_p^*a_{1}e^{i\omega_{p}t}).
%\end{eqnarray}
%In a rotating frame at the pump frequency $\omega_{L}$ and $\omega_{R}$, the system Hamiltonian reads as
%follows
\begin{eqnarray}
H=&&\sum_{k=1,2}\hbar\Delta_{k}a_{k}^\dagger a_{k}+\hbar\omega_{m}b^\dagger b-\sum_{k=1,2}\hbar g_k
a_k^\dagger a_k(b^\dagger+b)+i\hbar\sqrt{\kappa_{e,1}}E_{L}(a_{1}^\dagger -a_{1})\nonumber\\&&+i\hbar\sqrt{\kappa_{e,2}}E_{R}(a_{2}^\dagger -a_{2})+i\hbar\sqrt{\kappa_{e,1}}E_{p}(a_{1}^\dagger
e^{-i\delta t}-a_{1}e^{i\delta t}).
\end{eqnarray}
The first term describes the energy of the two optical cavity modes with resonance frequency $\omega_k(k=1,2)$, where
$a_k^\dagger$ $(a_k)$ is the creation (annihilation) operator of each cavity mode. $\Delta_1=\omega_1-\omega_L$ and $\Delta_2=\omega_2-\omega_R$ are the corresponding cavity-pump
field detunings. The second term gives the energy of the
mechanical mode with creation (annihilation) operator
$b^\dagger$ $(b)$, resonance frequency $\omega_{m}$ and effective
mass $m$. The third term is the radiation pressure coupling rate
$g_k=(\omega_k/L_k)\sqrt{\hbar/(2m\omega_m)}$, where $L_k$ is an effective length that depends on the cavity geometry. The last three terms represent the
input fields, where $E_L$, $E_R$, and $E_p$ are related to the power of the applied laser fields by $\left\vert
E_L\right\vert=\sqrt{2P_L\kappa_1/\hbar\omega_L}$, $\left\vert
E_R\right\vert=\sqrt{2P_R\kappa_2/\hbar\omega_R}$, and $\left\vert
E_p\right\vert=\sqrt{2P_p\kappa_1/\hbar\omega_p}$ ($\kappa_k$ the linewidth of the \emph{k}th cavity mode), respectively.
The total cavity linewidth $\kappa_k=\kappa_{i,k}+\kappa_{e,k}$, where $\kappa_{e,k}$ is the cavity decay rate due to coupling to an external photonic waveguide, as presented in the realistic two-mode optomechanical nanocavity \cite{Hill}.
$\delta=\omega_p-\omega_L$ is
the detuning between the probe filed and the left pump field.

Applying the Heisenberg equations of motion for operators $a_1$, $a_2$, and $Q$ which is defined as $Q=b^\dagger+b$ and introducing
the corresponding damping and noise terms \cite{Genes}, we derive the quantum Langevin equations as follows:
\begin{eqnarray}
&\dot{a_1}=-i(\Delta_1-g_1Q)a_1-\kappa_1 a_1+\sqrt{\kappa_{e,1}}(E_L+E_pe^{-i\delta t})+\sqrt{2\kappa_1}a_{in,1},\\
&\dot{a_2}=-i(\Delta_2-g_2Q)a_2-\kappa_2 a_2+\sqrt{\kappa_{e,2}}E_R+\sqrt{2\kappa_2}a_{in,2},\\
&\ddot{Q}+\gamma_m\dot{Q}+\omega_m^2 Q=2g_1\omega_ma_1^\dagger a_1+2g_2\omega_m a_2^\dagger a_2+\xi,
\end{eqnarray}
where $a_{in,1}$ and $a_{in,2}$ are the input vacuum noise operators with zero mean value, $\xi$ is the Brownian stochastic force with zero mean value \cite{Genes}.

Following standard methods from quantum optics,
we derive the steady-state solution to Eqs. (2)-(4) by setting all the time derivatives to zero. They are given by
\begin{eqnarray}
a_{s,1}=\frac{\sqrt{\kappa_{e,1}}E_L}{\kappa_1+i\Delta_1'}, a_{s,2}=\frac{\sqrt{\kappa_{e,2}}E_R}{\kappa_2+i\Delta_2'},
Q_s=\frac{2}{\omega_m}(g_1\left\vert a_{s,1}\right\vert^2+g_2\left\vert a_{s,2}\right\vert^2),
\end{eqnarray}
where $\Delta_1'=\Delta_1-g_1 Q_s$ and $\Delta_2'=\Delta_2-g_2 Q_s$ are the effective cavity detunings including radiation pressure effects.
We can rewrite each Heisenberg operator of Eqs. (2)-(4) as the sum of its steady-state mean value and a small fluctuation with zero mean value,
\begin{eqnarray}
a_1=a_{s,1}+\delta a_1, a_2=a_{s,2}+\delta a_2, Q=Q_s+\delta Q.
\end{eqnarray}Inserting these equations into the Langevin equations Eqs. (2)-(4) and assuming $\left\vert a_{s,1}\right\vert\gg1$ and $\left\vert a_{s,2}\right\vert\gg1$,
one can safely neglect the nonlinear terms $\delta a_1^\dagger\delta a_1$, $\delta a_2^\dagger\delta a_2$, $\delta a_1\delta Q$, and $\delta a_2\delta Q$. Since the drives are weak, but
classical coherent fields, we will identify all operators with their
expectation values, and drop the quantum and thermal noise terms
\cite{Weis}. Then the linearized Langevin equations can be written
as:
\begin{eqnarray}
&\left\langle\delta\dot{a_1}\right\rangle=-(\kappa_1+i\Delta_1)\left\langle\delta a_1\right\rangle+ig_1 Q_s\left\langle\delta a_1\right\rangle+ig_1 a_{s,1}\left\langle\delta Q\right\rangle+\sqrt{\kappa_{e,1}}E_p e^{-i\delta t},\\
&\left\langle\delta\dot{a_2}\right\rangle=-(\kappa_2+i\Delta_2)\left\langle\delta a_2\right\rangle+ig_2 Q_s\left\langle\delta a_2\right\rangle+ig_2 a_{s,2}\left\langle\delta Q\right\rangle,\\
&\langle\delta\ddot{Q}\rangle+\gamma_m\langle\delta\dot{Q}\rangle+\omega_m^2\langle\delta Q\rangle=2\omega_m g_1 a_{s,1}(\langle\delta a_1\rangle+\langle\delta a_1^\dagger\rangle)+2\omega_m g_2 a_{s,2}(\langle\delta a_2\rangle+\langle\delta a_2^\dagger\rangle).
\end{eqnarray}
In order to solve equations (7)-(9), we
make the ansatz \cite{Boyd} $\langle\delta a_1\rangle=a_{1+}e^{-i\delta
t}+a_{1-}e^{i\delta t}$, $\langle\delta a_2\rangle=a_{2+}e^{-i\delta
t}+a_{2-}e^{i\delta t}$, and $\langle\delta Q\rangle=Q_+e^{-i\delta
t}+Q_-e^{i\delta t}$. Upon substituting the above ansatz into Eqs. (7)-(9), we derive the following solution
\begin{eqnarray}
a_{1+}=\frac{\sqrt{\kappa_{e,1}}E_p}{\kappa_1+i\Delta_1'-i\delta}-\frac{1}{d(\delta)}\frac{ig_1^2 n_{1}\sqrt{\kappa_{e,1}}E_p}{(\kappa_1+i\Delta_1'-i\delta)^2},
\end{eqnarray}
where
\begin{eqnarray}d(\delta)=\sum_{k=1,2}\frac{2\Delta_k'g_k^2n_{k}}{(\kappa_k-i\delta)^2+\Delta_k'^2}
-\frac{\omega_m^2-\delta^2-i\delta\gamma_m}{\omega_m},
\end{eqnarray}
and $n_{k}=\left\vert a_{s,k}\right\vert^2.$
Here $n_{k}$, approximately equal to the number of
pump photons in each cavity, is determined by the following coupled equations
\begin{eqnarray}
n_{1}=\frac{\kappa_{e,1}E_L^2}{\kappa_1^2+\left[\Delta_1-2g_1/\omega_m(g_1n_{1}-g_2n_{2})\right]^2},\\
n_{2}=\frac{\kappa_{e,2}E_R^2}{\kappa_2^2+\left[\Delta_2-2g_2/\omega_m(g_1n_{1}-g_2n_{2})\right]^2}.
\end{eqnarray}

The output field can be obtained by employing the standard
input-output theory \cite{Gardiner}
$a_{out}(t)=a_{in}(t)-\sqrt{\kappa_e}a(t)$, where $a_{out}(t)$ is the
output field operator. Considering the output field of the left cavity, we have
\begin{eqnarray}
\left\langle a_{out}(t)\right\rangle&=&(E_L-\sqrt{\kappa_{e,1}}a_{s,1})e^{-i\omega_L t}+(E_p-\sqrt{\kappa_{e,1}}a_{1+})
e^{-i(\delta+\omega_L)t}-\sqrt{\kappa_{e,1}}a_{1-}e^{i(\delta-\omega_L)t}.
\end{eqnarray}
The transmission of the probe field, defined by the ratio of the
output and input field amplitudes at the probe frequency, is then
given by
\begin{eqnarray}
t(\omega_p)&=&\frac{E_p-\sqrt{\kappa_{e,1}}a_{1+}}{E_p}\nonumber\\&=&1-\left[\frac{\kappa_{e,1}}
{\kappa_1+i\Delta_1'-i\delta}-\frac{1}{d(\delta)}\frac{ig_1^2 n_{1}\kappa_{e,1}}{(\kappa_1+i\Delta_1'-i\delta)^2}\right].
\end{eqnarray}
The rapid phase dispersion $\phi=\arg[t(\omega_p)]$ of the transmitted probe laser beam leads to a group delay $\tau_{g}$ expressed as
\begin{eqnarray}
\tau_{g}=\left.\frac{d\phi}{d\omega_{p}}\right|_{\omega_p=\omega_{1}}.
\end{eqnarray}
Note that, if $E_R=0$ and $g_2=0$, the Eqs (10)-(16) lead to the well-known results for the single mode cavity optomechanical system, where electromagnetically induced transparency and slow light effect have been observed experimentally \cite{Weis,Naeini}. In what follows, we will investigate theoretically this phenomenon in the two-mode optomechanics we consider here.

\section{Numerical results and discussion}
To illustrate the numerical results, we choose a realistic two-mode cavity optomechanical
system to calculate the transmission spectrum of the probe field.
The parameters used are
\cite{Hill}: $\omega_1=2\pi\times205.3$ THz, $\omega_2=2\pi\times194.1$ THz, $\kappa_1=2\pi\times520$ MHz, $\kappa_2=1.73$ GHz, $\kappa_{e,1}=0.2\kappa_1$, $\kappa_{e,2}=0.42\kappa_2$, $\omega_m=2\pi\times4$ GHz, $Q_m=87\times10^3$,
 where $Q_{m}$ is the quality
factor of the nanomechanical resonator, and the damping rate $\gamma
_{m}$ is given by $\frac{\omega _{m}}{Q_m}.$ We can see that
$\omega_{1}>\kappa_1$ and $\omega_2>\kappa_2$, therefore the system operates in the
resolved-sideband regime also termed good-cavity limit necessary for the electromagnetically induced transparency.

Characterization of the optomechanical cavity can be performed by using two strong pump beams combined with a weak probe beam. With both pump beams detuned a mechanical frequency to the red of their respective cavity modes ($\Delta_1=\Delta_2=\omega_m$), a weak probe beam is then swept across the left cavity mode. The resulting transmission spectra of the probe beam as a function of the probe-cavity detuning $\Delta_p=\omega_p-\omega_1$ are plotted in Fig. 2, where $P_L=0, 0.1, 1$ and 10 $\mu$W, respectively, while the power of the right pump beam $P_R$ is kept equal to 0.1 $\mu$W. When $P_L= 0$ $\mu$W, there is a transmission dip in the center of the probe transmission spectrum, as shown in Fig. 2(a). However, as $P_L= 0.1$ $\mu$W, the broad cavity resonance splits into two dips and a narrow transparency window appears when the probe beam is resonant with the cavity frequency. As the pump power, and hence effective coupling strength $G_1=g_{1}\sqrt{n_1}$, increases further, so does the probe transmission at the cavity resonance. The width of the transparency window also increases and is given by the modified mechanical damping rate $\gamma_m^{\mathrm{eff}}\approx\gamma_m(1+C_1)$ \cite{Weis, Teufel1, Karuza}, where $C_1=G_1^2/\kappa_1\gamma_m$ is an equivalent optomechanical cooperativity parameter. This mechanically mediated electromagnetically induced transparency can be understood as a result of radiation pressure force oscillating at the beat frequency $\delta=\omega_p-\omega_L$ between the pump beam and the probe beam. If this driving force is close to the mechanical resonance frequency $\omega_m$, the vibrational mode is excited coherently, resulting in Stokes and anti-Stokes scattering of light from the strong pump field. If the cavity is driven on its red sideband, the highly off-resonant Stokes scattering is suppressed and only the anti-Stokes scattering builds up within the cavity. However, when the probe beam is resonant with the cavity, destructive interference with the anti-Stokes field suppresses its build-up and hence a transparency window appears in the probe transmission spectrum. Much as in atomic EIT, this effect causes an extremely steep dispersion for the transmitted probe photons, leading to a group delay.

Fig. 3 shows the magnitude and phase dispersion of the probe transmission as a function of probe-cavity detuning $\Delta_p$ with $\Delta_1=\Delta_2=\omega_m$ for $P_L=10$ $\mu$W and $P_R=0.1$ $\mu$W. It can be seen clearly that there is a transparency window combined with a steep positive phase dispersion at the cavity resonance, which will result in a tunable group delay of the transmitted probe beam. In addition, the delay in transmission is directly related to the advance on reflection through the bare cavity transmission contrast. To verify this, we plot the corresponding transmission group delay $\tau_g^{(T)}$ of the probe beam versus the left pump power $P_L$ with $\Delta_1=\Delta_2=\omega_m$ for $P_R=0.1$ $\mu$W in Fig. 4(a). As can be seen from the figure, the maximum transmission delay is $\tau_g^{(T)}\approx4.5$ ns. However, when the right pump beam beam is turned off, the transmission delay can be significantly increased, with a maximum delay 4 $\mu$s. In Fig. 4(c), we consider the effect of the external decay rate $\kappa_{e,1}$ on the group delay. If the external decay dominates the decay of the cavity, $\kappa_{e,1}=0.6\kappa$ for example, the maximum transmission group delay can be increased further. Moreover, the group delay of the reflected probe beam as a function of the power of the left pump beam is plotted in Fig. 4(d) where the delay is negative, representing group advance due to causality-preserving superluminal effects. Therefore, we can tune the group delay and advance of the probe beam by controlling the power of the pump beam.

We have then investigated the situation where the left cavity mode is driven by a blue-detuned pump beam while the right cavity mode is driven by a red-detuned pump beam, i.e., when $\Delta_1=-\omega_m$ and $\Delta_2=\omega_m$. In such a case, the probe beam is resonant with the Stokes field at the frequency $\omega_L-\omega_m$. Constructive interference between the probe beam and the Stokes field amplifies the transmitted probe beam in a narrow frequency window, which is the optomechanical analogue of electromagnetically induced amplification \cite{Naeini}. As shown in Fig. 5, where the probe transmission spectrum is plotted versus the probe-cavity detuning $\Delta_p$ for $P_L=P_R=0.1$ $\mu$W, an evident amplification of the transmitted probe beam can be seen with maximum transmission approximately equal to $140\%$. This kind of amplification have been observed in cavity optomechanics both in optical \cite{Karuza} and microwave \cite{Massel} domain. Here, we have demonstrated that amplification of the probe laser beam can be realized in a two-mode optomechanical system.

\section{conclusion}
In conclusion, we have demonstrated both electromagnetically induced transparency and amplification in a two-mode cavity optomechanics consisted of two optical cavity modes coupled to a common mechanical mode under different driving conditions.
Destructive interference between the probe beam and the anti-Stokes field leads to a transparency window in the probe transmission spectrum in conjunction with a steep positive phase dispersion, giving rise to the corresponding slow light effect. Our theoretical results show an optically tunable delay of 4 $\mu$s of the transmitted probe beam. Furthermore, constructive interference between the probe beam and the Stokes field can amplify the transmitted probe beam evidently.

\section{acknowledgments} The authors gratefully acknowledge support
from National Natural Science Foundation of China (No.11074088).

\newpage \centerline{\large{\bf Figure Captions}}

Figure 1 Schematic of a two-mode optomechanical system where two optical cavity modes, $a_1$ and $a_2$, are coupled to the same mechanical mode $b$. The left cavity is driven by a strong pump beam $E_L$ in the simultaneous presence of a weak probe beam $E_p$ while the right cavity is only driven by a pump beam $E_R$.

Figure 2 Probe transmission as a function of the probe-cavity detuning $\Delta_p=\omega_p-\omega_1$ for left pump power
$P_L$ equals to 0, 0.1, 1, and 10 $\mu$W, respectively. The right pump power is kept equal to 0.1 $\mu$W. Both the cavities are pumped on their respective red sidebands, i.e., $\Delta_1=\omega_m$ and $\Delta_2=\omega_m$.
Other parameters used are $\omega_1=2\pi\times205.3$ THz, $\omega_2=2\pi\times194.1$ THz, $\kappa_1=2\pi\times520$ MHz, $\kappa_2=1.73$ GHz, $\kappa_{e,1}=0.2\kappa_1$, $\kappa_{e,2}=0.42\kappa_2$, $\omega_m=2\pi\times4$ GHz, $Q_m=87\times10^3$.

Figure 3 (a) Magnitude and (b) phase of the transmitted probe beam versus probe-cavity detuning $\Delta_p$ for $P_L=10$ $\mu$W and $P_R=0.1$ $\mu$W. Other parameters are the same with figure 2.

Figure 4 Group delay $\tau_g$ of the (a)-(c) transmitted (d) reflected probe beam as a function of the left pump power $P_L$ with $\Delta_1=\omega_m$ and $\Delta_2=\omega_m$ considering the effects of $\kappa_{e,1}$ and $P_R$.
Other parameters are $\omega_1=2\pi\times205.3$ THz, $\omega_2=2\pi\times194.1$ THz, $\kappa_1=2\pi\times520$ MHz, $\kappa_2=1.73$ GHz, $\kappa_{e,2}=0.42\kappa_2$, $\omega_m=2\pi\times4$ GHz, $Q_m=87\times10^3$.

Figure 5 Probe transmission versus the probe-cavity detuning $\Delta_p$ for $P_L=P_R=0.1$ $\mu$W with the left cavity is pumped on its blue sideband while the right cavity is pumped on its red sideband, i.e., $\Delta_1=-\omega_m$ and $\Delta_2=\omega_m$. Optomechanically induced amplification appears in this case. Other parameters are the same with figure 2.

\clearpage
\begin{figure}
\includegraphics[width=12cm]{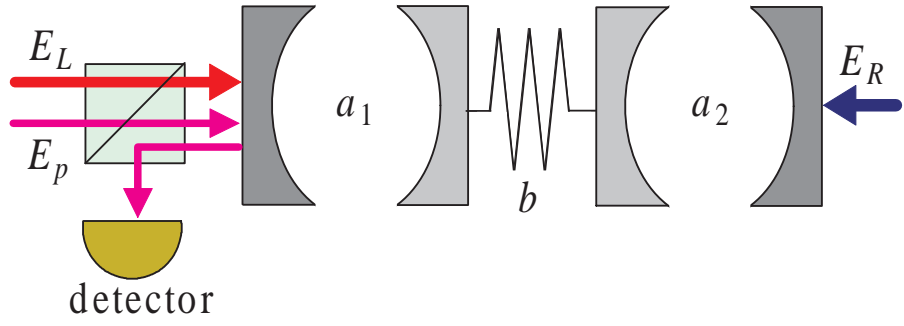}
\caption{Schematic of a two-mode optomechanical system where two optical cavity modes, $a_1$ and $a_2$, are coupled to the same mechanical mode $b$. The left cavity is driven by a strong pump beam $E_L$ in the simultaneous presence of a weak probe beam $E_p$ while the right cavity is only driven by a pump beam $E_R$.}
\end{figure}

\clearpage
\begin{figure}
\includegraphics[width=12cm]{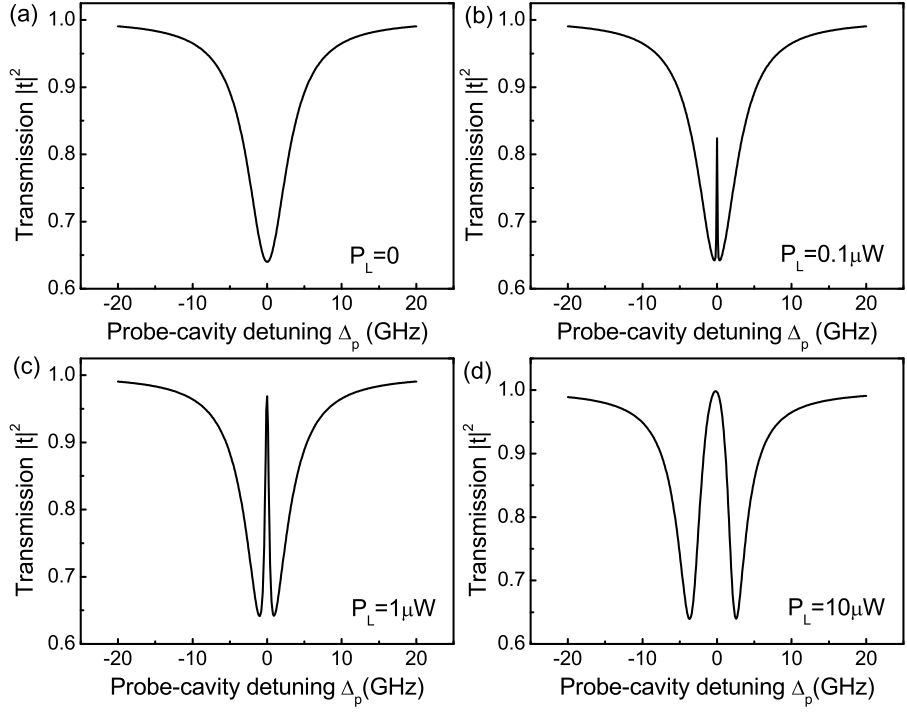}
\caption{Probe transmission as a function of the probe-cavity detuning $\Delta_p=\omega_p-\omega_1$ for left pump power
$P_L$ equals to 0, 0.1, 1, and 10 $\mu$W, respectively. The right pump power is kept equal to 0.1 $\mu$W. Both the cavities are pumped on their respective red sidebands, i.e., $\Delta_1=\omega_m$ and $\Delta_2=\omega_m$.
Other parameters used are $\omega_1=2\pi\times205.3$ THz, $\omega_2=2\pi\times194.1$ THz, $\kappa_1=2\pi\times520$ MHz, $\kappa_2=1.73$ GHz, $\kappa_{e,1}=0.2\kappa_1$, $\kappa_{e,2}=0.42\kappa_2$, $\omega_m=2\pi\times4$ GHz, $Q_m=87\times10^3$.}
\end{figure}

\clearpage
\begin{figure}
\centering
\includegraphics[width=12cm]{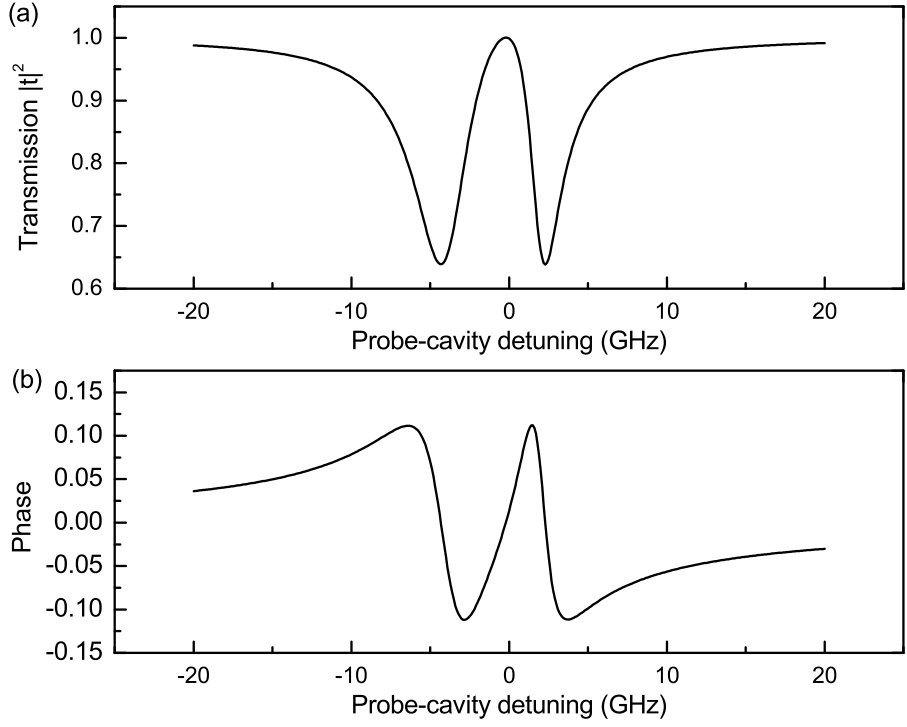}
\caption{(a) Magnitude and (b) phase of the transmitted probe beam versus probe-cavity detuning $\Delta_p$ for $P_L=10$ $\mu$W and $P_R=0.1$ $\mu$W. Other parameters are the same with figure 2.}
\end{figure}

\clearpage
\begin{figure}
\centering
\includegraphics[width=12cm]{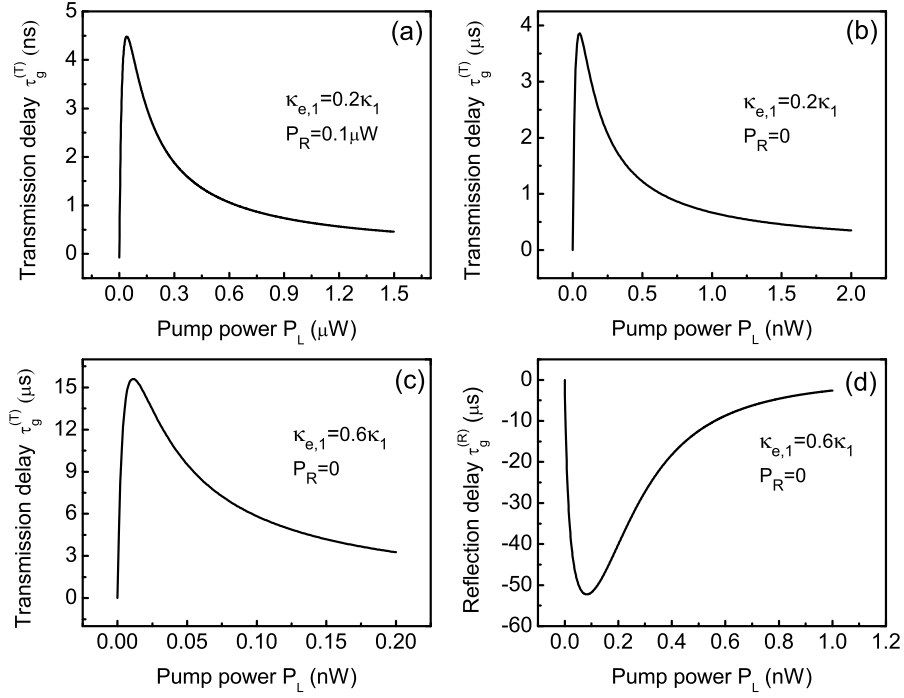}
\caption{Group delay $\tau_g$ of the (a)-(c) transmitted (d) reflected probe beam as a function of the left pump power $P_L$ with $\Delta_1=\omega_m$ and $\Delta_2=\omega_m$ considering the effects of $\kappa_{e,1}$ and $P_R$.
Other parameters are $\omega_1=2\pi\times205.3$ THz, $\omega_2=2\pi\times194.1$ THz, $\kappa_1=2\pi\times520$ MHz, $\kappa_2=1.73$ GHz, $\kappa_{e,2}=0.42\kappa_2$, $\omega_m=2\pi\times4$ GHz, $Q_m=87\times10^3$.}
\end{figure}

\clearpage
\begin{figure}
\centering
\includegraphics[width=12cm]{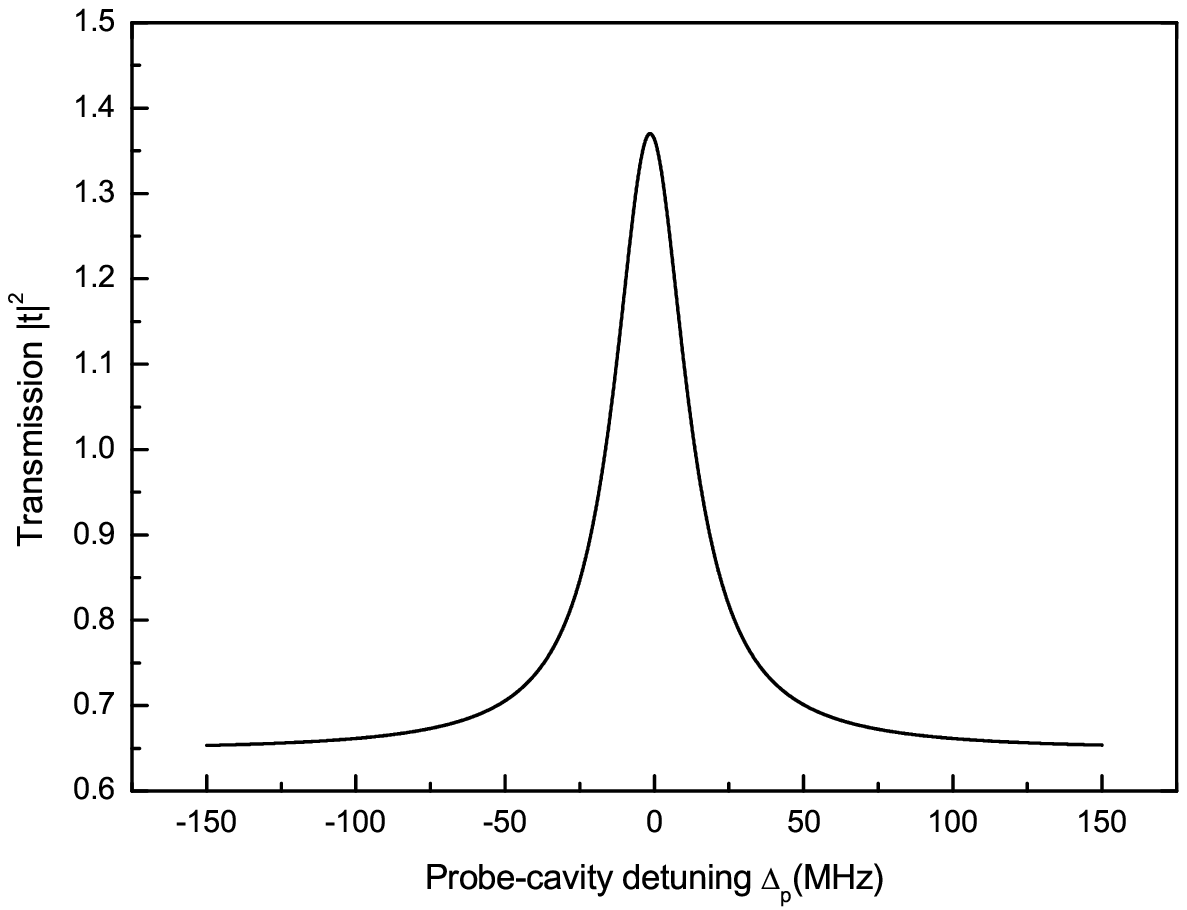}
\caption{Probe transmission versus the probe-cavity detuning $\Delta_p$ for $P_L=P_R=0.1$ $\mu$W with the left cavity is pumped on its blue sideband while the right cavity is pumped on its red sideband, i.e., $\Delta_1=-\omega_m$ and $\Delta_2=\omega_m$. Optomechanically induced amplification appears in this case. Other parameters are the same with figure 2.}
\end{figure}

\newpage
\centerline{\large{\bf References}}
\begin{thebibliography}{32}
\bibitem{Kippenberg1} T. J. Kippenberg and K. J. Vahala, Opt. Express \textbf{15,} 15172 (2007).

\bibitem{Kippenberg2} T. J. Kippenberg and K. J. Vahala, Science \textbf{321,} 1172 (2008).

\bibitem{Marquardt} F. Marquardt and S. M. Girvin, Physics \textbf{2,} 40 (2009).

\bibitem{Aspelmeyer} Aspelmeyer, P. Meystre, and K. Schwab, Phys. Today \textbf{65,} 29 (2012).

\bibitem{Brennecke} F. Brennecke, S. Ritter, T. Donner, and T. Esslinger, Science \textbf{322,} 235 (2008).

\bibitem{Rabl}P. Rabl, Phys. Rev. Lett. \textbf{107,} 063601 (2011).

\bibitem{Nunenkamp}A. Nunnenkamp, K. Borkje, and S. M. Girvin, Phys. Rev. Lett. \textbf{107,} 063602 (2011).

\bibitem{Teufel1}J. D. Teufel, D. Li, M. S. Allman,
K. Cicak, A. J. Sirois, J. D. Whittaker, and R. W. Simmonds, Nature (London)
\textbf{471}, 204 (2011).

\bibitem{Verhagen}E. Verhagen, S. Del¨¦glise, S. Weis, A. Schliesser, and T. J. Kippenberg, Nature (London) \textbf{482,} 63 (2012).
\bibitem{Teufel2}J. D. Teufel, T. Donner, D. Li, J. W. Harlow, M. S. Allman, K. Cicak, A. J. Sirois, J. D. Whittaker, K. W. Lehnert, and R. W. Simmonds, Nature (London) \textbf{475,} 359 (2011).

\bibitem{Chan}J. Chan, T. P. Alegre, A. H. Safavi-Naeini, J. T. Hill, A. Krause, S. Groeblacher, M. Aspelmeyer, and O. Painter, Nature (London) \textbf{478,} 89 (2011).

\bibitem{Groblacher}S. Gr\"{o}blacher, K. Hammerer, M. R. Vanner, and M.
Aspelmeyer, Nature (London) \textbf{460,} 724 (2009).

\bibitem{Agarwal} G. S. Agarwal and S. Huang, Phys. Rev. A \textbf{81}, 041803
(2010).

\bibitem{Weis} S. Weis, R. Rivi\`{e}re , S. Del\'{e}glise, E. Gavartin ,
O. Arcizet, A. Schliesser, and T. J. Kippenberg, Science
\textbf{330}, 1520 (2010).

\bibitem{Naeini} A. H. Safavi-Naeini, T. P. Mayer Alegre, J. Chan, M. Eichenfield, M. Winger, Q. Lin, J. T.Hill, D.
E. Chang, and O. Painter, Nature (London) \textbf{472}, 69 (2011).

\bibitem{Karuza} M. Karuza, C. Biancofiore, C. Molinelli, M. Galassi, R. Natali, P. Tombesi, G. Di Giuseppe, D. Vitali, arXiv:1209.1352 (2012).

\bibitem{Fleischhauer}M. Fleischhauer, A. Imamoglu, and J. P. Marangos, Rev. Mod. Phys. \textbf{77,} 633 (2005).

\bibitem{Hau}L.V. Hau, S. E. Harris, Z. Dutton, and C. H. Behroozi, Nature (London) \textbf{397,} 594 (1999).

\bibitem{Chang}D. E. Chang, A. H. Safavi-Naeini, M. Hafezi, and O. Painter, New J. Phys. \textbf{13,} 023003 (2011).

\bibitem{Boller}K.-J. Boller, A. Imamo\v{g}lu, and S. E. Harris, Phys. Rev. Lett. \textbf{66,} 2593 (1991).

\bibitem{Phillips} M. C. Phillips, H. Wang, I. Rumyantsev, N. H. Kwong, R. Takayama, and R. Binder, Phys. Rev. Lett. \textbf{91,} 183602 (2003).

\bibitem{Liu}N. Liu, L. Langguth, T. Weiss, J. K\"{a}stel, M. Fleischhauer, T. Pfau, and H. Giessen, Nat. Mater. \textbf{8,} 758 (2009).

\bibitem{Santori}C. Santori, P. Tamarat, P. Neumann, J. Wrachtrup, D. Fattal, R. G. Beausoleil, J. Rabeau, P. Olivero, A. D. Greentree, S. Prawer, F. Jelezko, and P. Hemmer, Phys. Rev. Lett. \textbf{97,} 247401 (2006).

\bibitem{Zhou}X. Zhou, F. Hocke, A. Schliesser, A. Marx, H. Huebl, R. Gross, T. J. Kippenberg, arXiv:1206.6052 (2012).

\bibitem{Ludwig}M. Ludwig, A. H. Safavi-Naeini, O. Painter, and F. Marquardt, Phys. Rev. Lett. \textbf{109,} 063601 (2012).

\bibitem{Komar}P. K\'{o}m\'{a}r, S. D. Bennett, K. Stannigel, S. J. M. Habraken, P. Rabl, P. Zoller, M. D. Lukin, arXiv:1210.4039 (2012).

\bibitem{Stannigel}K. Stannigel, P. Komar, S. J. M. Habraken, S. D. Bennett, M. D. Lukin, P. Zoller, and P. Rabl, Phys. Rev. Lett. \textbf{109,} 013603 (2012).

\bibitem{Qu}K. Qu and G. S. Agarwal, arXiv:1210.4067 (2012).

\bibitem{Hill} J. T. Hill, A. H. Safavi-Naeini, J. Chan, O. Painter, Nat. Commun. \textbf{3,} 1196 (2012).

\bibitem{Dong} C. Dong, V. Fiore, M. C. Kuzyk, L. Tian, H. Wang, arXiv:1205.2360 (2012).

\bibitem{Massel} F. Massel, T. T. Heikkil\"{a}, J.-M. Pirkkalainen, S. U. Cho, H. Saloniemi, P. Hakonen, M. A. Sillanp\"{a}\"{a}, Nature \textbf{480}, 351-154 (2011).

\bibitem{Genes}C. Genes, D. Vitali, P. Tombesi, S. Gigan, and M. Aspelmeyer, Phys. Rev. A \textbf{77}, 033804 (2008).

\bibitem{Boyd} R. W. Boyd  {\it Nonlinear Optics} (San Diego, CA:
Academic) (2008).

\bibitem{Gardiner} C. W. Gardiner and P. Zoller {\it Quantum Noise} (Springer) (2004).

\end{thebibliography}
\end{document}